\documentclass{article}
\usepackage{amsfonts}
\usepackage{amsmath}

\input{tcilatex}

\begin{document}

\date{}
\title{Some Algebraic Geometry Aspects of Gravitational Theories with
Covariant and Contravariant Connections and Metrics (GTCCCM) and Possible
Applications to Theories with Extra Dimensions}
\author{Bogdan G. Dimitrov \\
{\small \textit{Bogoliubov Laboratory for Theoretical Physics}}\\
[-1.mm] {\small \textit{Joint Institute for Nuclear Research, Dubna 141 980,
Russia}}\\
[-1,mm] {\small \textit{email: bogdan@theor.jinr.ru}}}
\maketitle

\begin{abstract}
\ On the base of the distinction between covariant and contravariant metric
tensor components, an approach from algebraic geometry will be proposed,
aimed at finding new solutions of the Einstein's equations both in GTCCCM
and in standard gravity theory, if these equations are treated as algebraic
equations.

\ \ \ As a partial case, some physical applications of the approach have
been considered in reference to theories with extra dimensions. The s.c.
"length function" $l(x)$ has been introduced and has been found as a
solution of quasilinear differential equations in partial derivatives for
two different cases, corresponding to "compactification + rescaling" and
"rescaling + compactification" of the type I low-energy string theory
action. New (although complicated) relations between the parameters in the
action have been found, valid also for the standard approach in theories
with extra dimensions. 

\ \ \ \ 
\end{abstract}

\section{Introduction and Statement of the Problem}

Inhomogeneous cosmological models have been intensively studied in the past
in reference to colliding gravitational \bigskip waves [1] or singularity
structure and generalizations of the Bondi - Tolman and Eardley-Liang-Sachs
metrics [2, 3]. In these models the inhomogeneous metric is assumed to be of
the form [2] 
\begin{equation}
ds^{2}=dt^{2}-e^{2\alpha (t,r,y,z)}dr^{2}-e^{2\beta (t,r,y,z)}(dy^{2}+dz^{2})
\tag{1.1}
\end{equation}%
(or with $r\rightarrow z$ and $z\rightarrow x$) and it is called the
Szafron-Szekeres metric [4-7]. In [7], after an integration of one of the
components - $G_{1}^{0}$ of the Einstein's equations, a solution in terms of
an elliptic function is obtained.

This is an important point since valuable cosmological characteristics for
observational cosmology such as the Hubble's constant $H(t)=\frac{\overset{.}%
{R}(t)}{R(t)}$ and the deceleration parameter $q=-\frac{\overset{..}{R}%
(t)R(t)}{\overset{.}{R}^{2}(t)}$ may be expressed in terms of the Jacobi's
theta function and of the Weierstrass elliptic function respectively [8]. In
the same paper, the expression for the metric of the anisotropic
cosmological model in the Szafron-Szekeres approach has been obtained in
terms of the Weierstrass elliptic function after reducing the component $%
G_{1}^{0}$ of the Einstein's equations [7, 8] to the nonlinear differential
equation \ 
\begin{equation}
\left( \frac{\partial \Phi }{\partial t}\right) ^{2}=-K(z)+2M(z)\Phi ^{-1}+%
\frac{1}{3}\Lambda \Phi ^{2}\text{ \ .}  \tag{1.2}
\end{equation}%
Then if one sets up 
\begin{equation}
g_{2}=\frac{K^{2}(z)}{12}\text{ \ \ ; \ \ \ }g_{3}=\frac{1}{216}K^{3}(z)-%
\frac{1}{12}\Lambda M^{2}(z)\text{ \ \ ,}  \tag{1.3}
\end{equation}

the resulting two-dimensional cubic algebraic equation 
\begin{equation}
y^{2}=4x^{3}-g_{2}x-g_{3}\text{ \ \ \ }  \tag{1.4}
\end{equation}

according to the standard algebraic geometry prescription (see [9] for a
contemporary introduction into algebraic geometry)\ can be parametrized as 
\begin{equation}
x=\rho (z)\text{ \ \ },\text{ \ }y=\rho ^{^{\prime }}(z)\text{ \ \ \ ,} 
\tag{1.5}
\end{equation}%
where $\rho (z)$ is the well-known Weierstrass elliptic function 
\begin{equation}
\rho (z)=\frac{1}{z^{2}}+\sum\limits_{\omega }\left[ \frac{1}{(z-\omega )^{2}%
}-\frac{1}{\omega ^{2}}\right]  \tag{1.6}
\end{equation}%
and the summation is over the poles in the complex plane. However, it is
important to remind the standard definition of the elliptic curve (1.5),
according to which the points $z$ on the complex plane, factorized by the
lattice $\Lambda =\{m\omega _{1}+n\omega _{2}\mid m,n\in Z;$ $\omega
_{1},\omega _{2}\in C,Im\frac{\omega _{1}}{\omega _{2}}>0\},$ are mapped via
the mapping $f:$ $C/\Lambda \rightarrow CP^{2}$ onto the points $(x,y)=(\rho
(z),\rho ^{^{\prime }}(z))$ of the two \textbf{dimensional complex
projective space }$CP^{2}$, which satisfy the curve (1.5). This definition
in fact imposes an important restriction on the functions $g_{2}$ and $g_{3}$%
, namely they should be equal to the s.c. "Eisenstein series" 
\begin{equation}
g_{2}=60\sum\limits_{\omega \subset \Gamma }\frac{1}{\omega ^{4}};\text{ \ \ 
}g_{3}=140\sum\limits_{\omega \subset \Gamma }\frac{1}{\omega ^{6}}\text{ \
\ .}  \tag{1.7}
\end{equation}%
In other words, if the parametrization (1.5) is possible also for the
nonlinear differential equation (1.2), then the Eisenstein series imply
additional restrictions on the functions in the right-hand side of (1.3),
which has not been taken into account in the paper [8]. However, there is
another reasoning if one admits that the Weierstrass function and its
derivatives can parametrize a cubic algebraic equation of the type (1.3)
with non-constant coefficient functions (depending on the complex variable $%
z $) - to a certain extent this problem has been analysed in [10].

\section{\protect\bigskip Fundamentals of the New Algebraic Geometry
Approach in Gravity Theory}

The main goal of the present paper and of the preceeding ones [10, 11] is to
propose a new algebraic geometry approach for finding new solutions of the
Einstein's equations by representing these equations in an algebraic form.
The approach also is based on the gravitational theory with covariant and
contravariant metrics and connections (GTCCMC) [12]. This theory makes a
clear distinction between covariant $g_{ij}$ and contravariant metric tensor
components $\widetilde{g}^{is}$, which means that $\widetilde{g}^{is}$
should not be considered to be the inverse ones to the covariant components $%
g_{ij}$, consequently $\widetilde{g}^{is}g_{im}\equiv f_{m}^{s}(\mathbf{x})$%
. It is elementary to prove [11,12] that if the components of $f_{m}^{s}(%
\mathbf{x})$ are considered to be functions, then two connections should be
introduced - one connection $\Gamma _{\alpha \beta }^{\gamma }$ for the case
of the parallel transport of covariant basic vectors $\nabla _{e_{\beta
}}e_{\alpha }=\Gamma _{\alpha \beta }^{\gamma }$ $e_{\gamma }$ and a \textit{%
separate} connection $P_{\alpha \beta }^{\gamma }$ for the contravariant
basic vector $e^{\gamma }$, the defining equation for which is $\nabla
_{e_{\beta }}e^{\alpha }=P_{\gamma \beta }^{\alpha }$ $e^{\gamma }$.
However, if $f_{m}^{s}(\mathbf{x})$ are considered to be tensor components,
then by simple covariant differentiation of $\widetilde{g}^{is}g_{im}\equiv
f_{m}^{s}(\mathbf{x})$ it can be proved that it is not obligatory to
introduce a second connection $P_{\gamma \beta }^{\alpha }$, i.e. it can be
assumed that $P_{\gamma \beta }^{\alpha }=-\Gamma _{\gamma \beta }^{\alpha }$%
, as it is the case in the standard Einsteinian theory of gravity. This
shall further be exploited, when assuming that the contravariant metric
tensor can be represented in the form of the factorized product $\widetilde{g%
}^{ij}=dX^{i}dX^{j}$, where the differentials $dX^{i}$ remain in the tangent
space $T_{X}$ of the generalized coordinates $%
X^{i}=X^{i}(x_{1},x_{2},....,x_{n})$, defined on the initially given
manifold. Also, the existence of different from $g^{ij}$ contravariant
metric tensor components $\widetilde{g}^{ij}$ means that another connection 
\begin{equation}
\widetilde{\Gamma }_{kl}^{s}\equiv \widetilde{g}^{is}\Gamma _{i;kl}=%
\widetilde{g}^{is}g_{im}\Gamma _{kl}^{m}=\frac{1}{2}\widetilde{g}%
^{is}(g_{ik,l}+g_{il,k}-g_{kl,i})\text{ \ ,}  \tag{2.1}
\end{equation}%
not consistent with the initial metric $g_{ij}$, can be introduced. By
substituting $\widetilde{\Gamma }_{kl}^{s}$ in the expression for the
"tilda" Ricci tensor $\widetilde{R}_{ij}$ and requiring the equality of the
"tilda" scalar curvature $\widetilde{R}$ with the usual one $R$, i.e. $%
\widetilde{R}=R$, one can obtain the s.c. "cubic algebraic equation for
reparametrization invariance of the gravitational Lagrangian" [10] 
\begin{equation}
dX^{i}dX^{l}\left( p\Gamma _{il}^{r}g_{kr}dX^{k}-\Gamma
_{ik}^{r}g_{lr}d^{2}X^{k}-\Gamma _{l(i}^{r}g_{k)r}d^{2}X^{k}\right)
-dX^{i}dX^{l}R_{il}=0\text{ \ \ \ .}  \tag{2.2}
\end{equation}%
In the same way, assuming the contravariant metric tensor components to be
equal to the "tilda" ones, the Einstein's equations in vacuum were derived.
They can be derived also in the general case for arbitrary $\widetilde{g}%
^{ij}$, when the assumption about the factorized representation $\widetilde{g%
}^{ij}=dX^{i}dX^{j}$ is no longer implemented: 
\begin{equation*}
0=\widetilde{R}_{ij}-\frac{1}{2}g_{ij}\widetilde{R}=
\end{equation*}%
\begin{equation*}
=\widetilde{g}^{lr}(\Gamma _{r;i[j}),_{l]}+\widetilde{g}_{,[l}^{lr}\Gamma
_{r;ij]}+\widetilde{g}^{lr}\widetilde{g}^{ms}(\Gamma _{r;ij}\Gamma
_{s;lm}-\Gamma _{s;il}\Gamma _{r;km})-
\end{equation*}%
\begin{equation*}
-\frac{1}{2}g_{ij}\widetilde{g}^{m[k}\widetilde{g}_{,l}^{l]s}\Gamma
_{mk}^{r}g_{rs}-\frac{1}{2}g_{ij}\widetilde{g}^{m[k}\widetilde{g}%
^{l]s}\left( \Gamma _{mk}^{r}g_{rs}\right) _{,l}-
\end{equation*}%
\begin{equation}
-\frac{1}{2}g_{ij}\widetilde{g}^{nk}\widetilde{g}^{ls}\widetilde{g}%
^{mr}g_{pr}g_{qs}\left( \Gamma _{nk}^{q}\Gamma _{lm}^{p}-\Gamma
_{nl}^{p}\Gamma _{km}^{q}\right) \text{ \ .}  \tag{2.3}
\end{equation}

Interestingly, this "algebraic" system of the Einstein's equations can be
considered as well as a system of \textit{fifth - degree} algebraic
equations with respect to the covariant variables and also as a system of
third-rank equations (with non-constant coefficient functions) with respect
to the contravariant variables and their derivatives.The mathematical
treatment of fifth - degree equations is known since the time of Felix
Klein's famous monograph [13], published in 1884. A way for resolution of
such equations has been developed also in [9, 14, 15].

\section{Application to Theories with Extra Dimensions. Tensor Length Scale.}

In view of the application of the algebraic geometry approach in theories
with extra dimensions, the cubic algebraic equation for reparametrization
invariance of the gravitational Lagrangian in the general case (for
arbitrary $\widetilde{g}^{ij}$) (eq. (5.2) in [11]) 
\begin{equation*}
\widetilde{g}^{i[k}\widetilde{g}_{,l}^{l]s}\Gamma _{ik}^{r}g_{rs}+\widetilde{%
g}^{i[k}\widetilde{g}^{l]s}\left( \Gamma _{ik}^{r}g_{rs}\right) _{,l}+
\end{equation*}%
\begin{equation}
+\widetilde{g}^{ik}\widetilde{g}^{ls}\widetilde{g}^{mr}g_{pr}g_{qs}\left(
\Gamma _{ik}^{q}\Gamma _{lm}^{p}-\Gamma _{il}^{p}\Gamma _{km}^{q}\right) -R=0%
\text{ \ \ \ \ .}  \tag{3.1}
\end{equation}%
can be represented in the form [16] 
\begin{equation*}
\varepsilon \frac{\partial l}{\partial y}+\frac{h^{^{\prime }}}{8k}%
e^{2k\varepsilon y}\left( \frac{\partial l}{\partial x_{1}}-\frac{\partial l%
}{\partial x_{2}}-\frac{\partial l}{\partial x_{3}}-\frac{\partial l}{%
\partial x_{4}}\right) =
\end{equation*}%
\begin{equation}
=(2k-\varepsilon \frac{h^{^{\prime }}}{h})(l^{3}-l)\text{ \ \ }  \tag{3.2}
\end{equation}%
for the special case of the metric of a $4D$ flat Minkowski space, embedded
in a five-dimensional space 
\begin{equation}
ds^{2}=e^{-2k\epsilon y}\eta _{\mu \nu }dx^{\mu }dx^{\nu }+h(y)dy^{2}\text{
\ \ \ \ \ \ ,}  \tag{3.3}
\end{equation}%
where $\eta _{\mu \nu }=(+,-,-,-)$ and $\epsilon =\pm 1$. Note that the
metric (3.3) will be with a negative constant (scalar) curvature if
additionally one can set up $h(y)=1$. The obtained equation (3.2) is no
longer treated as an algebraic one, but as a quasilinear differential
equation in partial derivatives with respect to the function $l(x)$, which
was introduced in [10,16] and has been called "the length function". The key
idea for its introduction is that the contraction of the covariant metric
tensor $g_{ij}$ with the contravariant one $\widetilde{g}^{jk}=dX^{j}dX^{k}$
gives exactly (when $i=k$) the length interval 
\begin{equation}
l=ds^{2}=g_{ij}dX^{j}dX^{i}\text{ \ \ .}  \tag{3.4}
\end{equation}%
Then naturally, for $i\neq k$ the contraction will give a tensor function $%
l_{i}^{k}=g_{ij}dX^{j}dX^{k}$ (since it is assumed also that the
differentials $dX^{j}$ in the tangent space are vectors - in the general
case this might not be true), which can be interpreted as a
\textquotedblright tensor\textquotedblright\ length scale for the different
directions. Thus in deriving the quasilinear differential equation (3.2) it
had been assumed that $\widetilde{g}^{is}g_{im}\equiv f_{m}^{s}(\mathbf{x}%
):=l_{i}^{k}=l\delta _{i}^{k}$. In principle, it can be accepted that a
"hint" for the necessity for introduction of such a "tensor scale" is
contained in Witten's paper [17], where it was remarked that
\textquotedblright the problem might be ameliorated by considering an
anisotropic Calabi - Yau with a scale $\sqrt{\alpha ^{^{\prime }}}$\ in $d$\
directions and $\frac{1}{M_{GUT}}$\ in $(6-d)$\ directions\textquotedblright
.

The quasilinear differential equation (3.2) has the following solution for
the square of the length function

\begin{equation}
l^{2}=\frac{1}{1-const.\text{ }e^{24\text{ }k\text{ }\varepsilon \text{ }y}}%
\text{ \ \ \ \ \ \ }\varepsilon =\pm 1\text{ \ \ ,}  \tag{3.5}
\end{equation}%
From a physical point of view it is interesting to note that the
\textquotedblright scale function\textquotedblright\ will indeed be equal to
one (i.e. we have the usual gravitational theory with $\widetilde{g}%
^{ij}=g^{ij}$ and $l=1$) for $\varepsilon =-1$ and $y\rightarrow \infty $
(the s.c. infinite extra dimensions).

\section{Compactification in Low Energy Type I Ten-Dimensional String Theory
- A Brief Review of the Standard Approach}

The standard approach is based on the low - energy action of type I string
theory in ten dimensions [18, 19, 20] 
\begin{equation}
S=\int d^{10}x\left( \frac{m_{s}^{8}}{(2\pi )^{7}\lambda ^{2}}R+\frac{1}{4}%
\frac{m_{s}^{6}}{(2\pi )^{7}\lambda }F^{2}+...\right) \text{ }=\int
d^{4}xV_{6}(......)\text{\ \ \ \ ,}  \tag{4.1}
\end{equation}%
where $\lambda \sim \exp (\Phi )$ is the string coupling , $m_{s}$ is the
string scale, which we can identify with $m_{grav.}$. Compactifying to $4$
dimensions on a manifold of volume $V_{6}$, one can identify the resulting
coefficients in front of the $R$ and $\frac{1}{4}F^{2}$ terms with $%
M_{(4)}^{2}$ and $\frac{1}{g_{4}^{2}}$, from where one obtains [18] 
\begin{equation}
M_{(4)}^{2}=\frac{(2\pi )^{7}}{V_{6}m_{s}^{4}g_{4}^{2}}\text{ \ \ \ ; \ \ \ }%
\lambda =\frac{g_{4}^{2}V_{6}m_{s}^{6}}{(2\pi )^{7}}\text{ \ \ \ . } 
\tag{4.2}
\end{equation}
Note that although one of the key assumptions in extra-dimensional theories
is that the volume $V_{6}$ is very large (and thus the scale for the
four-dimensional gravitational constant $M_{(4)}^{2}$ is considerably
lowered), there is a large indeterminacy, since the exact value of $V_{6}$
is not known.

\section{"Compactification + Rescaling" and "Rescaling + Compactification" -
A Proposed New Approach}

The essence of the proposed new approach is that the operation of
compactification is "supplemented" by the additional operation of
"rescaling" of the contravariant metric tensor components in the sense,
which was clarified in Section 2. One can discern two cases:

1st case - "compactification + rescaling". One starts from the
\textquotedblright unrescaled\textquotedblright\ ten - dimensional action
(4.1), then performs a compactification to a five - dimensional manifold and
afterwards\textbf{\ }a transition to the usual \textquotedblright
rescaled\textquotedblright\ scalar quantities $\widetilde{R}$ and $%
\widetilde{F}^{2}$. Then it is required that the \textquotedblright
unrescaled\textquotedblright\ ten - dimensional effective action (4.1) (i.e.
the L. H. S. of (2.1)) is equivalent to the five - dimensional effective
action after compactification, but in terms of the rescaled quantities $%
\widetilde{R}$ and $\widetilde{F}^{2}$ in the R. H. S of (4.1). This can be
expressed as follows \ 
\begin{equation*}
S=\int d^{10}x\left( \frac{m_{s}^{8}}{(2\pi )^{7}\lambda ^{2}}R+\frac{1}{4}%
\frac{m_{s}^{6}}{(2\pi )^{7}\lambda }F^{2}\right) =\int d^{4}xV_{6}\left(
....\right) =
\end{equation*}%
\begin{equation}
=\int d^{4}x\left( M_{(4)}^{2}\widetilde{R}+\frac{1}{4}\frac{1}{g_{4}^{2}}%
\widetilde{F}^{2}\right) \text{ \ \ \ \ .}  \tag{5.1}
\end{equation}%
Identifying the expressions in front of the \textquotedblright
unrescaled\textquotedblright\ scalar quantities $F^{2}$ and $R\,$, one
obtains an algebraic relation and a differential equation in partial
derivatives. Combining them, the following differential equation is obtained
[16]: 
\begin{equation*}
\frac{(2\pi )^{7}l^{3}}{m_{s}^{4}V_{6}g_{4}^{4}}\left( \frac{\partial l}{%
\partial x^{C}}g^{AB}\Gamma _{AB}^{C}-\frac{\partial l}{\partial x^{B}}%
g^{AB}\Gamma _{AC}^{C}\right) +\frac{RP^{2}}{l^{3}(P-NR)^{2}}\left[ \frac{%
(2\pi )^{7}l^{2}}{m_{s}^{4}V_{6}g_{4}^{4}}-M_{4}^{2}\right] +
\end{equation*}%
\begin{equation}
+\frac{(2\pi )^{7}l(l-1)}{m_{s}^{4}V_{6}g_{4}^{4}}g^{AB}\left( \Gamma
_{AB}^{C}\Gamma _{CD}^{D}-\Gamma _{AC}^{D}\Gamma _{BD}^{C}\right) =0\text{ \
\ \ \ . }  \tag{5.2}
\end{equation}%
2nd case - "rescaling + compactification". In this case, first a rescaling
of the contravariant metric components is performed, which means that one
starts from the ten-dimensional string theory action with "rescaled"
components 
\begin{equation*}
S=\int d^{10}x\left( \frac{m_{s}^{8}}{(2\pi )^{7}\lambda ^{2}}\widetilde{R}+%
\frac{1}{4}\frac{m_{s}^{6}}{(2\pi )^{7}\lambda }\widetilde{F}^{2}\right)
=\int d^{4}xV_{6}\left( ....\right) =
\end{equation*}%
\begin{equation}
=\int d^{5}x\left( M_{(4)}^{2}R+\frac{1}{4}\frac{1}{g_{4}^{2}}F^{2}\right) 
\text{ \ \ \ \ ,}  \tag{5.3}
\end{equation}%
and after that the compactification is realized, resulting again in the
right-hand side of the standard $4D$ action (5.1), but this time with
unrescaled components. In an analogous way, the following algebraic relation
can be obtained [16] 
\begin{equation}
\left[ \frac{(2\pi )^{7}}{V_{6}m_{s}^{4}g_{4}^{4}}-M_{(4)}^{2}\right] R=%
\frac{(2\pi )^{7}(l^{2}-1)}{2m_{s}^{4}V_{4}l^{2}g_{4}^{2}}g^{AC}g^{BD}(....)%
\text{ \ \ ,}  \tag{5.4}
\end{equation}%
where for brevity the brackets denote $%
(....):=(g_{AD,BC}+g_{BC,AD}-g_{AC,BD}-g_{BD,AC})$. For $l=1$, as expected,
we obtain the usual relation for $M_{(5)}^{2}$ as in (4.2). Now introducing
the notation 
\begin{equation}
\beta \equiv \left[ \frac{(2\pi )^{7}}{V_{6}m_{s}^{4}g_{4}^{4}}-M_{(4)}^{2}%
\right] m_{s}^{4}V_{6}\frac{2}{(2\pi )^{7}}\text{ \ ,}  \tag{5.5}
\end{equation}%
and assuming a small deviation from the relation $M_{(4)}^{2}=\frac{(2\pi
)^{7}}{V_{6}m_{s}^{4}g_{4}^{2}}$ , i.e. $\beta \ll 1$,one can express the
length scale $l(x)$ from the algebraic relation (5.4) as 
\begin{equation}
l^{2}=\frac{1}{1-\beta \frac{R}{g^{AC}g^{BD}(...)}}\approx 1+\beta \frac{R}{%
g^{AC}g^{BD}(...)}\text{ \ \ \ .}  \tag{5.6}
\end{equation}%
\textit{Consequently the deviation from the \textquotedblright
standard\textquotedblright\ length scale }$l=1$\textit{\ in the case of a
gravitational theory with }$l\neq 1$\textit{\ in the case of small }$\beta $%
\textit{\ shall be proportional to the ratio }$\frac{R}{g^{AC}g^{BD}(...)}$,
which for the assumed metric (3.3) with $h(y)=1$ is exactly equal to $\frac{1%
}{4}$ and thus, this constant number will not affect the smallness of the
number $\beta \frac{R}{g^{AC}g^{BD}(...)}$.\textit{The above result may have
also an important physical meaning - the zero value of the number }$\beta $%
\textit{\ (which signifies the fulfillment of the relation }$M_{(4)}^{2}=%
\frac{(2\pi )^{7}}{V_{6}m_{s}^{4}g_{4}^{2}}$\textit{) can be related to the
usual length scale }$l=1$\textit{\ in gravity theory. Conversely, any
deviations from the relation (4.2) }may be attributed to deviations from the
usual scale $l=1$. Note also that the assumption $\beta \ll 1$ poses some
restriction on the number $m_{s}^{4}V_{6}\frac{2}{(2\pi )^{7}}$ in (5.5) -
it should not be too large.

The differential equation in partial derivatives for the scale function $%
l(x) $ for this case ("rescaling + compactification") is different from the
preceeding one (5.2) 
\begin{equation*}
\left[ \frac{(2\pi )^{7}}{m_{s}^{4}V_{6}g_{4}^{4}l^{2}}-M_{4}^{2}\right] R+%
\frac{(2\pi )^{7}}{m_{s}^{4}V_{6}g_{4}^{4}}\frac{(l-1)}{l^{2}}g^{AB}\left(
\Gamma _{AB}^{C}\Gamma _{CD}^{D}-\Gamma _{AC}^{D}\Gamma _{BD}^{C}\right) +
\end{equation*}%
\begin{equation}
+\frac{(2\pi )^{7}}{m_{s}^{4}V_{6}g_{4}^{4}}\frac{1}{l^{3}}\left[ \frac{%
\partial l}{\partial x^{C}}g^{AB}\Gamma _{AB}^{C}-\frac{\partial l}{\partial
x^{B}}g^{AB}\Gamma _{AC}^{C}\right] =0\text{ \ \ .}  \tag{5.7}
\end{equation}

3rd case - simultaneous fulfillment of \textit{\textquotedblright rescaling
+ compactification\textquotedblright\ and \textquotedblright
compactification + rescaling\textquotedblright }. In the formal sense, this
may mean that it does not matter in what sequence the two operations are
performed, i.e. the process of compactification is accompanied by rescaling.
From the simultaneous fulfillment of the two differential equations (5.2)
and (5.7) one obtains the following cubic algebraic equation with respect to 
$l^{2}$: 
\begin{equation}
M_{4}^{2}l^{6}-\frac{(2\pi )^{7}}{m_{s}^{4}V_{6}g_{4}^{4}}l^{4}+\frac{(2\pi
)^{7}}{m_{s}^{4}V_{6}g_{4}^{4}}\frac{P^{2}}{(P-NR)^{2}}l^{2}-M_{4}^{2}\frac{%
P^{2}}{(P-NR)^{2}}=0\text{.}  \tag{5.8}
\end{equation}%
The length function $l(x)$ can be an imaginary one for the case of the
imaginary Lobachevsky space [12], realized by all the straight lines outside
the absolute cone (on which the scalar product is zero, i.e. $[x,x]=0$- the
length may may take imaginary values in the interval $[0,\frac{\pi \text{ }i%
}{2k}]$,where $k$ is the Lobachevsky constant). Further we shall assume that 
$l(x)$ is a real function, which will mean that the roots $l^{2}(x)$ of the
cubic equation have to be positive. From this requirement one obtains the
following inequualities, relating the parameters in the low energy type I
string theory action 
\begin{equation}
p^{2}=\frac{b^{2}}{2}+\frac{a^{3}}{27}-b\sqrt{\frac{b^{2}}{2}+\frac{a^{3}}{27%
}}>\left[ \frac{a_{1}+6a}{18}+\frac{a_{1}}{18}\sqrt{a_{1}^{2}+12a}\right]
^{3}  \tag{5.9}
\end{equation}%
or for 
\begin{equation}
p^{2}=\frac{b^{2}}{2}+\frac{a^{3}}{27}-b\sqrt{\frac{b^{2}}{2}+\frac{a^{3}}{27%
}}<\left[ \frac{a_{1}+6a}{18}-\frac{a_{1}}{18}\sqrt{a_{1}^{2}+12a}\right]
^{3}\text{ \ \ ,}  \tag{5.10}
\end{equation}%
where 
\begin{equation}
a\equiv a_{2}-\frac{a_{1}^{2}}{3}\text{ \ \ ; \ \ \ }b\equiv 2\frac{a_{1}^{3}%
}{27}-\frac{a_{1}a_{2}}{3}+a_{3}\text{ \ \ \ ,}  \tag{5.11}
\end{equation}%
\begin{equation}
a_{1}\equiv -\frac{(2\pi )^{7}}{M_{4}^{2}m_{s}^{4}V_{6}g_{4}^{2}}\text{ \ \
\ ; \ \ \ \ }a_{2}\equiv \frac{(2\pi )^{7}}{%
M_{4}^{2}m_{s}^{4}V_{6}g_{4}^{2}Q^{2}}\text{ \ \ ; \ \ }a_{3}\equiv -\frac{%
g_{4}^{2}}{Q^{2}}\text{ \ ,}  \tag{5.12}
\end{equation}%
\begin{equation}
Q\equiv \frac{g^{AC}g_{BD}(2\pi )^{7}g_{4}^{4}(....)}{\left[
g^{AC}g^{BD}(...)(2\pi )^{7}g_{4}^{4}-2R\left( (2\pi
)^{7}-M_{4}^{2}V_{6}m_{s}^{4}g_{4}^{4}\right) \right] }\text{ \ .} 
\tag{5.13}
\end{equation}%
\textit{The last two inequalities (5.9) - (5.10) are new (although rather
complicated) inequalities between the parameters in the low-energy type I
string theory action, which cannot be obtained in the framework of the usual
gravity theory. Also, it is impoortant to stress that since the scale
function }$l(x)$ does not enter in them, they are valid for the standard
gravity theory approach in theories with extra dimensions.

\end{document}